\documentclass[aps,prc,twocolumn,showpacs,superscriptaddress,floatfix,raggedbottom]{revtex4-1}
\usepackage{amsmath, amsthm, amssymb}
\usepackage{graphics}
\usepackage{graphicx}
\usepackage{epsfig}
\usepackage{float}
\usepackage{multirow}   
\usepackage{array}
\begin{document}
\title{Competition between direct and sequential two-neutron transfers in the $^{18}$O + $^{28}$Si collision at 84 MeV}
\author{E. N. Cardozo}
\author{J. Lubian}
\author{R.Linares}
\affiliation{Instituto de F\'isica, Universidade Federal Fluminense, 24210-340, Niter\'oi, Rio de Janeiro, Brazil}
\author{F. Cappuzzello}
\affiliation{Istituto Nazionale di Fisica Nucleare, Laboratori Nazionali del Sud, I-95125 Catania, Italy}
\affiliation{Dipartimento di Fisica e Astronomia, Universit\`a di Catania, I-95125 Catania, Italy}
\author{D. Carbone}
\author{M. Cavallaro}
\affiliation{Istituto Nazionale di Fisica Nucleare, Laboratori Nazionali del Sud, I-95125 Catania, Italy}
\author{J.L.Ferreira}
\affiliation{Instituto de F\'isica, Universidade Federal Fluminense, 24210-340, Niter\'oi, Rio de Janeiro, Brazil}
\author{A. Gargano}
\affiliation{Istituto Nazionale di Fisica Nucleare, Sezione di Napoli, Napoli, Italy}
\author{B.Paes}
\affiliation{Instituto de F\'isica, Universidade Federal Fluminense, 24210-340, Niter\'oi, Rio de Janeiro, Brazil}
\author{G. Santagati}
\affiliation{Istituto Nazionale di Fisica Nucleare, Laboratori Nazionali del Sud, I-95125 Catania, Italy}

\date{\today}

\begin{abstract} 

In this work we study the simultaneous and sequential two-neutron transfer mechanisms to the $^{28}$Si nucleus induced by (t,p) and ($^{18}$O, $^{16}$O) reactions. New experimental cross sections for the $^{28}$Si($^{18}$O,$^{16}$O)$^{30}$Si reaction at 84 MeV are also presented. Direct reaction calculations are carried out within the Exact Finite Range Coupled Reaction Channel, for the simultaneous transfer of the two-neutron cluster, and the second order Distorted Wave Born Approximation, for the sequential transfer. Two different models are considered to describe the two-neutron cluster. The spectroscopic information was obtained from shell model calculation with {\it psdmod} interaction for the target overlaps where the 1p$_{3/2}$, 1p$_{1/2}$, 1d$_{3/2}$, 1d$_{5/2}$ and 2s$_{1/2}$ orbitals are included as valence sub-space. We show that simultaneous and sequential two-neutron transfer are competing mechanisms for the population of the ground state in $^{30}$Si. A systematic analysis of the two-neutron transfer induced by the ($^{18}\textnormal{O},^{16}\textnormal{O}$) indicates that static deformation of target nuclei impacts on the two-neutron transfer mechanism.

\end{abstract}

\pacs{}

\maketitle

\section{Introduction}
\label{Intro}

The response of atomic nuclei to external probes exhibits interesting multi facets features. Depending on the case, single particle, cluster and collective degrees of freedom may be distinguished in the many body problem. Single particle configurations are mainly determined by the nuclear mean field, while cluster and shape deformed configurations are connected with nucleon-nucleon correlations beyond the mean field. In particular, nucleon-paired configurations are connected to short-range correlations while states with deformed shapes require long-range correlations. 

Pairing is in the foreground in nuclei with two neutrons or two protons outside a doubly-magic core. The two-neutron transfer reaction is a suitable tool to assess pairing correlations above the Fermi level in many nuclear systems. In the past, most of the two-neutron transfer measurements were conducted primarily with (t,p) reactions \cite{MFM78,BaH73,ACH86,ABS80,CBF81} but the use of triton beams is nowadays restricted also due to radiation protection. The ($^{18}$O,$^{16}$O) reaction is also an effective probe to access two-neutron pairing configurations, since the two-neutron system is pre-formed in the $^{18}$O nucleus and the beam production is straightforward. The systematic investigation using the ($^{18}$O,$^{16}$O) transfer reactions at the same bombarding energies have been published recently \cite{CCC11,CCB13,CCB14,CBB14,PSR15,ECL16,ELL17,CFC17,PSM17,AGC18}. In Ref.\cite{CCB13}, the experimental cross sections for one- and two-neutron transfer reactions in $^{12,13}$C($^{18}$O,$^{17}$O)$^{13,14}$C and $^{12}$C($^{18}$O,$^{16}$O)$^{14}$C were reproduced by direct reaction calculations for the first time, without requiring arbitrary scaling  factors. Such theoretical calculations indicate the dominance of simultaneous two-neutron transfer (e.g. $^{12}$C $\rightarrow ^{14}$C) over the sequential one (e.g. $^{12}$C $\rightarrow ^{13}$C $\rightarrow ^{14}$C). The same conclusions have been obtained in Refs.\cite{ECL16,CFC17} for $^{16}$O and $^{13}$C target nuclei. In the light of such results, the pairing-like configuration of the final states favors the simultaneous transfer mechanism.

In Ref. \cite{PSM17}, the two-neutron transfer induced by the ($^{18}$O,$^{16}$O) reaction to the $^{64}$Ni nucleus has been studied. Again, the results indicate that simultaneous transfer is dominant for the population of the ground state in $^{66}$Ni. However, for the first excited state (2$^{+}_{1}$), the sequential mechanism competes with the simultaneous one. The 2$^{+}_{1}$ state in $^{66}$Ni is characterized by a collective component that smears the pairing correlation of the two transferred neutrons and suppresses the simultaneous transfer mechanism. 

The two-neutron transfer seems to be sensitive to the interplay between short-range (pairing) and long-range (collective) interactions. Understanding the effects of collectivity of the final states on two-neutron transfer mechanisms requires further studies. Following this line, transfer to the $^{28}$Si nucleus seems to be a good benchmark since the ground state is deformed and low-lying excited states can be interpreted within the rotor model.

In the past the two-neutron transfer to $^{28}\textnormal{Si}$ has been studied by (t,p) \cite{ACH86,BaH73},  and ($^{18}\textnormal{O},^{16}\textnormal{O}$) \cite{MFG79} reactions. In Ref.~\cite{ACH86}, the $^{28}$Si(t,p)$^{30}$Si reaction was studied at 18 MeV incident energy. The theoretical calculations were performed within the distorted wave Born approximation (DWBA) using the shell model (\textit{sd} space) to describe low-lying states in the $^{28,30}\textnormal{Si}$ isotopes. The agreement between experimental and theoretical cross sections required a normalization of the calculation in order to reproduce the absolute value of the experimental data. The same reaction was also studied at 10.5 and 12.1 MeV  \cite{BaH73} incident energy. However, the thickness of the target was not known with sufficient accuracy to determine a reliable absolute cross section. 


The angular distribution of multi-nucleon transfer reactions induced by $^{18}$O on $^{28}$Si was studied in  Ref.~\cite{MFG79} at 56 MeV incident energy. Optical model calculations for elastic and inelastic scattering, DWBA for one-step processes and coupled channel Born approximation (CCBA) for inelastic excitations were performed. The shape of the angular distribution was described reasonably well adopting the cluster approximation. Once again, normalization factors were necessary to reproduce the order of magnitude of the experimental cross sections. 


In this work, we present new experimental data for the two-neutron transfer in the $^{28}$Si($^{18}$O,$^{16}$O)$^{30}$Si reaction at 84 MeV and revisit the experimental data reported in Refs.~\cite{MFG79,ACH86} for the ($^{18}$O,$^{16}$O) reaction at $56~\textnormal{MeV}$ and the (t,p) reaction at $18~\textnormal{MeV}$. This set of experimental data is used to assess spectroscopic parameters and optical potentials that enters into the direct reaction calculations. 

This paper is organized as follows: the experimental details and the theoretical analysis are discussed in sections~\ref{exp} and~\ref{theor}, respectively, and the conclusions are given in section~\ref{conc}.

\section{\label{exp}Experimental details}

The measurements were performed at the Istituto Nazionale di Fisica Nucleare - Laboratori Nazionali del Sud, Catania, Italy. The 84 MeV $^{18}$O$^{6+}$ beam was delivered by the Tandem accelerator. A $^{28}$Si (136  $\mu$g/cm$^{2}$ thickness) self-supporting foil was used as target. The $^{16}$O$^{8+}$ ejectiles from the reaction were momentum analyzed by the MAGNEX spectrometer \cite{LCC07,LCC08,CCC07,CaC16} set in the full acceptance mode ($\Omega \sim 50$ msr). Parameters of the final trajectory (i.e. vertical and horizontal positions and incident angles) were measured by the focal plane detector that also allows for particle identification \cite{CCC10}. Trajectory reconstruction of $^{16}$O ejectiles was performed by solving the equation of motion for each particle to obtain scattering parameters at the target, according to procedures described in Refs. \cite{LCC07,LCC08,CAB14,Car15}.

The reaction was measured at two angular settings, with the spectrometer optical axis centered at $\theta_{lab}$ = 8$^\circ$ and $10\circ$. Due to the large angular acceptance of the spectrometer, these angular settings correspond to a total covered angular range of $4\circ < \theta_{\textnormal{lab}} < 15\circ$ in the laboratory framework, with an overlap of $\sim 8\circ$ between the two settings.

\begin{figure}[tb!]
\centering
\graphicspath{{figuras/}}
\includegraphics[width=0.45\textwidth]{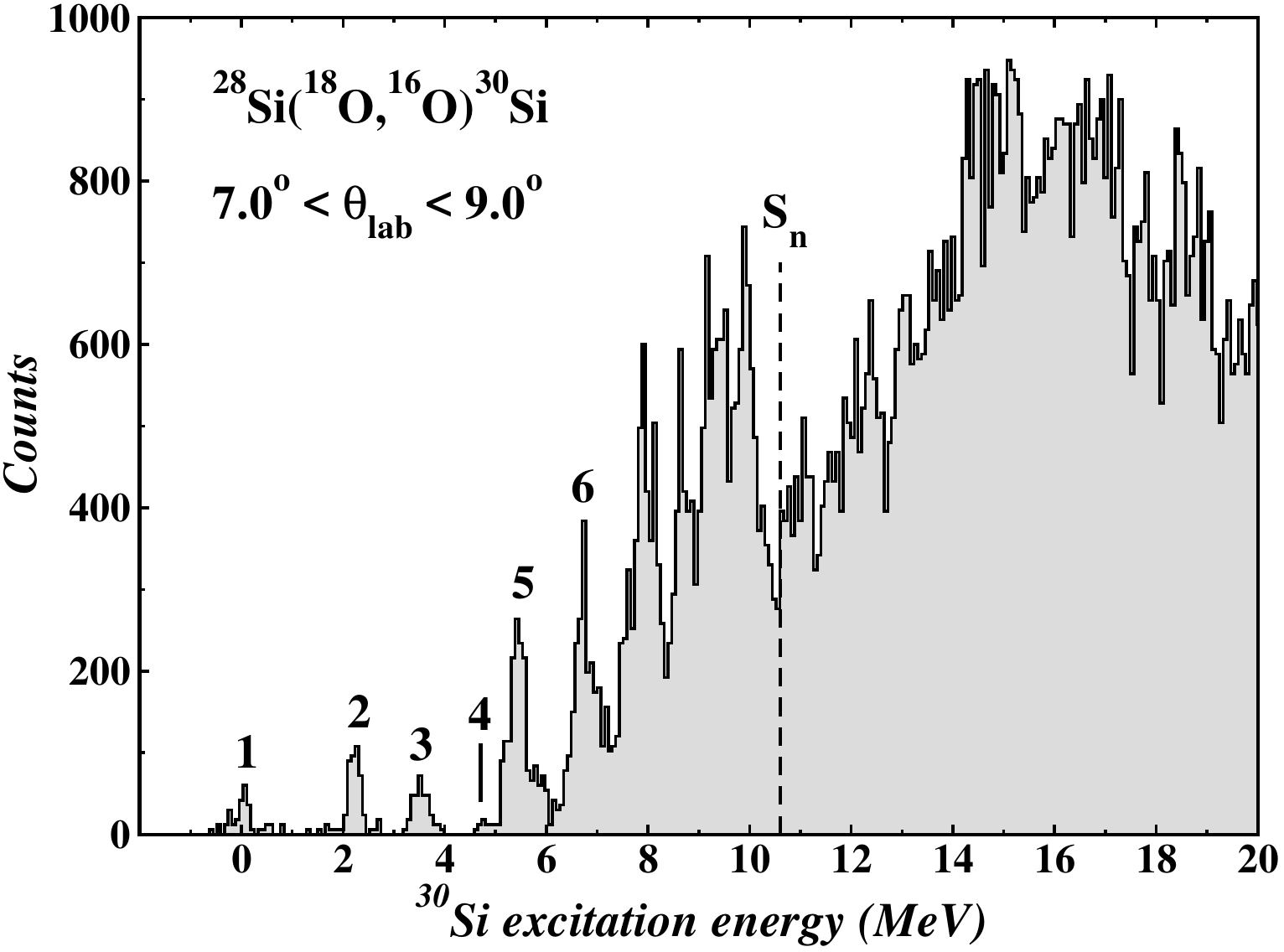}
\caption{Excitation energy spectrum of $^{30}$Si populated by the $^{28}$Si($^{18}$O,$^{16}$O) reaction at $E_{\textnormal{lab}} = 84 \textnormal{MeV}$. Low-lying states in $^{30}$Si are labeled with numbers (see Table \ref{table:ListOfStates}) and the one-neutron threshold energy (S$_{\textnormal{n}}$) is indicated with a dashed line.} 
\label{spectrum}
\end{figure}

The $^{30}\textnormal{Si}$ excitation energy spectrum, relative to ground to ground states $Q$-value of the reaction ($6.89\textnormal{ MeV}$), is shown in Fig.~\ref{spectrum}. The energy resolution, estimated from the full width half maximum of the ground state peak, is about 250 keV and allows for a clear identification of the ground and the $2_{1}^{+}$ (2.25 MeV) states of $^{30}$Si. The optimum Q-value region corresponds to about 16 MeV in excitation energy while the optimum angular momentum transfer is L$_{opt}$ = 4. Several peaks are observed below the neutron separation energy ($\textnormal{S}_{\textnormal{n}} = 10.6\textnormal{ MeV}$). The continuous shape, observed at excitation energies higher than the neutron separation energy, contains a contribution from the three-body kinematics connected to the one-neutron emission. Possible peaks due to contamination of the $^{28}\textnormal{Si}$ foil, usually $^{12}\textnormal{C}$ and $^{16}\textnormal{O}$ incorporated during fabrication and handling, could interfere at excitation energies higher than 7.1 MeV. Nevertheless they are not clearly observed in the spectrum. 

\begin{table} [H]
\caption{List of low-lying states (label and excited energies) identified in the energy spectrum of Fig.~\ref{spectrum}. The spin and parity assignment are taken from Ref.~\cite{ACH86}. Experimental angle-integrated cross section are extracted from $5.5\circ< \theta_{\textnormal{lab}} < 16.0\circ$.  } 
\centering
\begin{tabular}{c c c c} 
\hline
 \textbf{label} & \textbf{exc. energy} & J$^{\pi}$ & \textbf{cross section} \\
                &     (MeV)           &            & (mb) \\ 
\hline
\hline
                       
    1     & g.s.        & 0$^{+}_{1}$              & 0.17  \\ \cline{1-4}
    2     & 2.24        & 2$^{+}_{1}$              & 0.22  \\ \cline{1-4}        

          & 3.50        & 2$^{+}_{2}$              &       \\ 
    3     & 3.77        & 1$^{+}_{1}$              & 0.30  \\ 
          & 3.79        & 0$^{+}_{2}$              &       \\ \hline

          & 4.81        & 2$^{+}_{3}$              &       \\ 
    4     & 4.83        & 3$^{+}_{1}$              & 0.09  \\  \hline
          
          & 5.28        & 4$^{+}_{1}$              &       \\ 
          & 5.37        & 0$^{+}_{3}$              &       \\           
    5     & 5.49        & 3$^{-}_{1}$              & 0.87  \\ 
          & 5.61        & 2$^{+}_{4}$              &       \\           
          & 5.95        & 4$^{+}_{1}$              &       \\ \hline     
          
          & 6.50        & 4$^{-}_{1}$              &       \\ 
          & 6.54        & 2$^{+}_{5}$              &       \\ 
    6     & 6.64        & 2$^{-}_{1}$/0$^{+}_{4}$  & 0.79  \\ 
          & 6.74        & 1$^{-}_{1}$              &       \\ 
          & 6.87        & 3$^{+}_{2}$              &       \\ 
          & 6.91        & 0$^{+}_{5}$              &       \\ \hline            
\end{tabular}
\label{table:ListOfStates}
\end{table} 

Some low-lying states in $^{30}\textnormal{Si}$ have been identified by comparison with the results of the (t,p) reaction at 18 MeV \cite{ACH86}. The peaks labeled with 3 to 6 correspond to a set of states, listed in Table~\ref{table:ListOfStates}, with indication of the excited energy, spin, parity and the angle-integrated cross sections. The $2^{+}_{3}$ and $3^{+}_{1}$ states (label 4) are weakly populated. The set of states labeled as 5 is a combination of 5 states ($4^{+}_{1}$,  $0^{+}_{3}$, $3^{-}_{1}$, $2^{+}_{4}$ and $4^{+}_{1}$), among which the $3^{-}_{1}$ state is the most intense, according to Refs.~\cite{MFG79,ACH86}. A similar situation appears in the set of states 6, in which the $1^{-}_{1}$ state is the most intense, according to Refs. \cite{ACH86,MFG79}.

Angular distributions of absolute cross sections for the ground and the $2_{1}^{+}$ states in $^{30}$Si are shown in Fig. \ref{ic+seq+cluster}. They were obtained individually whereas the peak at around 3.5 MeV (label 3) was assumed to have contribution from the unresolved $2_{2}^{+}$, $1_{1}^{+}$ and $0_{2}^{+}$ states. Cross sections were derived considering the counting statistics within an angular resolution of $0.3\circ$. A scale error in the cross section of 10$\%$, coming from systematics uncertainties in the target thickness and beam integration by the Faraday cup, is common to all the angular distribution points and it is not included in the error bars. These correspond to other sources of uncertainty, such as the solid angle determination and counting statistics.

\section{\label{theor}Theoretical Analysis}

Direct reaction calculations for two-neutron transfers were performed using prior exact finite range within the coupled channel Born approximation (CCBA) and coupled reaction channel (CRC) frameworks using the FRESCO code \cite{Tho88}. Non-orthogonality corrections and full complex remnant terms were considered in the coupled channel equations. The S\~ao Paulo double folding potential (SPP) \cite{CPH97} was used for the real and imaginary parts of the optical potential. As usual, the imaginary strength factor was set to 0.6 in the initial partition to account for missing couplings to continuum states, not explicitly considered \cite{PLO09}. In the exit partitions, the imaginary part was scaled by a larger factor (0.78) to avoid double counting the effect of continuum states. Optical model calculations using these coefficients provide a good description of the elastic scattering cross section for many systems in a wide energy interval \cite{GCG06,PLO12,OCC13}.

The adopted deformation parameter for the collective states in the $^{28}$Si target nucleus is $\beta_2 = 0.407$, taken from Ref.\cite{RNT01}. The single-particle and cluster wave functions used in the matrix elements calculations were generated by Woods-Saxon potentials, whose depth was varied in order to reproduce the experimental separation energies for one- and two-neutron, respectively. The reduced radii and diffuseness parameters were set to 1.26 fm and 0.65 fm, respectively, for $^{28}$Si and $^{29}$Si cores.  For the $^{16}$O and $^{17}$O cores the reduced radii and diffuseness were 1.26 fm and 0.70 fm, as recently done for the $^{64}$Ni($^{18}$O,$^{16}$O)$^{66}$Ni reaction\cite{PSM17}.

\subsection{Two-neutron transfer reaction}

Simultaneous (or one-step) and sequential (or two-step) transfer mechanisms were considered separately. In the first case, the two particles are transferred simultaneously and the wave functions used in the CRC calculations were obtained within two schemes: i) the cluster and ii) the independent coordinates. In the sequential mechanism, the two neutrons are transferred one by one, through the intermediate partition ($^{17}$O + $^{29}$Si). Excitations of low-lying states in the entrance partition are considered within the two-step CCBA formalism. 

In the cluster model, the relative motion between the two transferred neutrons is frozen and separated from the core. In this sense, the two-neutron transfer process is equivalent to the single-particle with the anzats that the correct quantum numbers of the cluster have to be considered. In this model, the intrinsic spin of the two-neutron cluster can be $S = 0$ (anti-parallel) or $S = 1$ (parallel). The wave function of the cluster is defined by the following quantum numbers: the orbital angular momentum $L$ relative to the core, the principal quantum number $N$ and the transferred angular momentum $J$. $N$ and $L$ can be determined from the conservation of the total number of quanta in the transformation of the wave function of two independent neutrons in orbits $(n_i, l_i)$ $(i = 1, 2)$ into a cluster with internal state $(n, l)$ \cite{Mos59}: $\sum_{i=1}^{2}2(n_i-1)+l_i=2(N-1)+L+2(n - 1) + l$. The cluster model is called extreme cluster approximation when we consider only $S = 0$ anti-parallel configuration. In this case, the intrinsic cluster wave function has the quantum numbers $n = 1$ and $l = 0$ so that the cluster is in the 1s internal state. The spectroscopic amplitudes for both projectile and target overlaps were set to 1.0, as usually found in the literature. 

For the independent coordinate (IC) and sequential (Seq) models, calculations are performed using microscopic information obtained by the shell model calculations. The spectroscopic amplitudes were calculated by the NuShellX code \cite{Rae08}. For silicon isotopes, the \textit{psdpn} model space, with the effective phenomenological interaction \textit{psdmod} \cite{UC11}, was considered. That model space assumes $^4$He as a closed core and valence  neutrons and protons in the 1p$_{3/2}$, 1p$_{1/2}$, 1d$_{3/2}$, 1d$_{5/2}$, and 2s$_{1/2}$ orbits. This interaction was generated from a modification of the \textbf{psdwbt} one \cite{UC11}. The hamiltonian is similar to the one used by Warburton, Brown, and Millener (WBM) to describe the excited states of $^{16}$O \cite{WBM92}, where they used a two-body hamiltonian that gives a global fit to \textit{p-sd}-shell nuclei. 

\begin{figure}[H]
\centering
\graphicspath{{figuras/}}
	\includegraphics[width=0.40\textwidth]{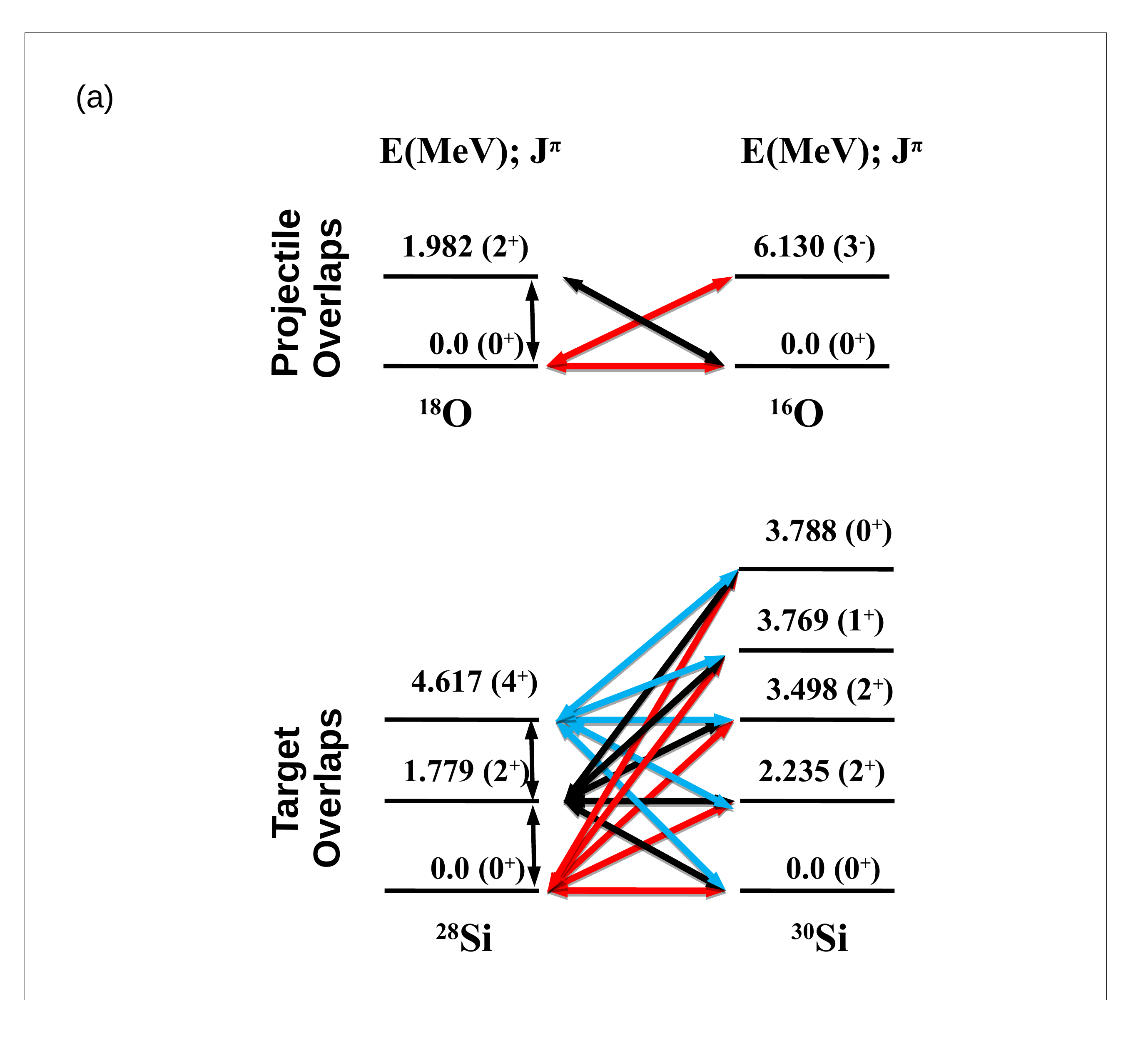}
	\label{fig:config-cluster}
\quad 
	\includegraphics[width=0.40\textwidth]{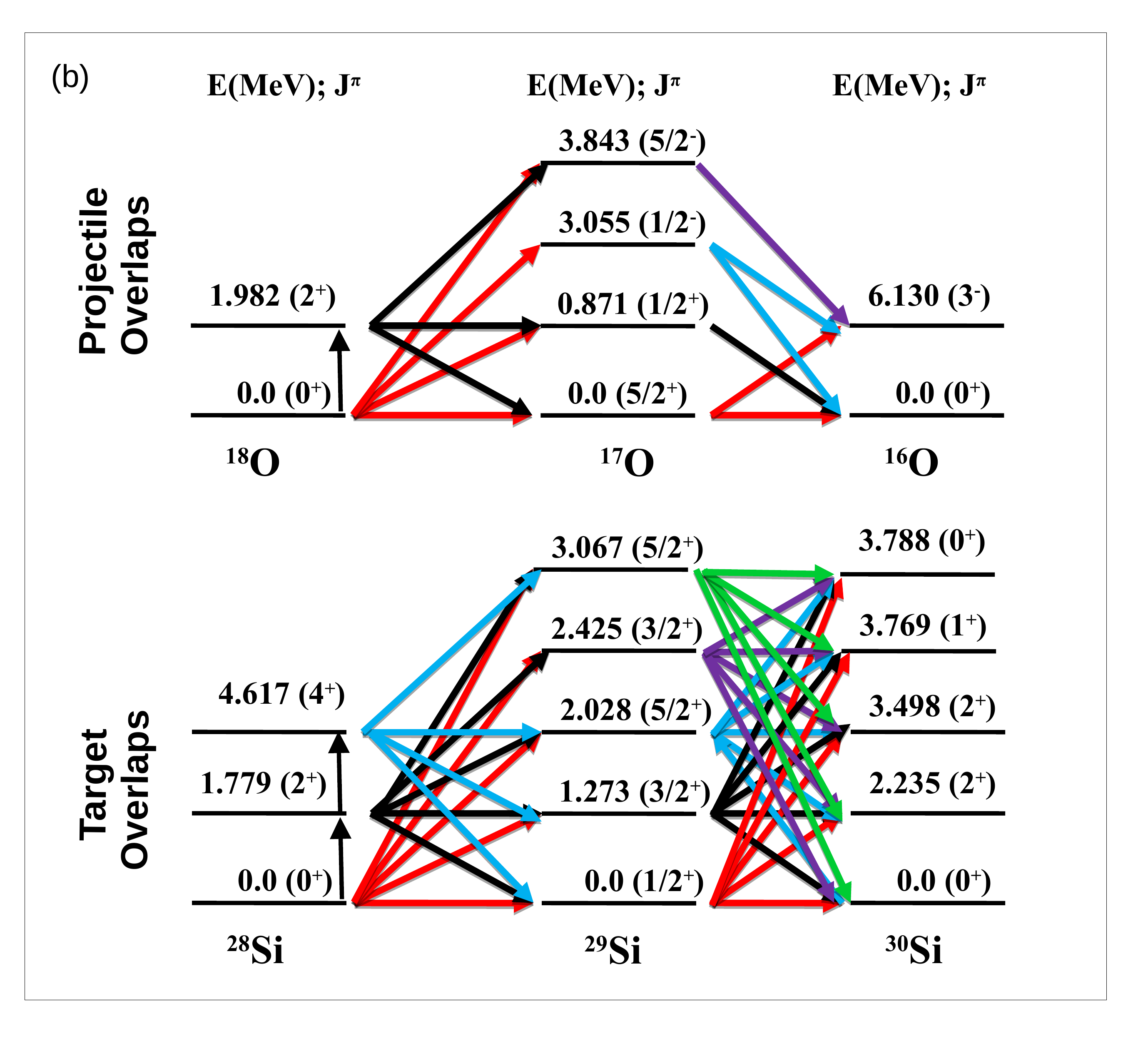}
	\label{fig:config-seq}
	\caption{Coupling schemes of the projectile and target overlaps used in the direct (a) and sequential (b) two-neutron transfer reaction calculations.}
\label{fig2}
\end{figure}

The adopted model space allows us to successfully reproduce spin, parity and relative energies of low-lying states in $^{28}$Si, $^{29}$Si and $^{30}$Si isotopes with differences between experimental and shell model results on excited energies better than 350 KeV, as shown in Table~\ref{energia}. The two-neutron spectroscopic amplitudes for simultaneous transfer within the IC scheme are listed in Tables \ref{table-ic}-\ref{table-ic3} and for sequential mechanism in Tables \ref{seq} and \ref{seq1}. The couplings and level schemes of the nuclei are sketched in Figure \ref{fig2}. Coupling schemes and the spectroscopic amplitudes for $^{16}$O, $^{17}$O and $^{18}$O are the same of Refs.\cite{ECL16,CFC17,PSM17}, obtained from shell model calculations using the\textit{ zbm} interaction \cite{ZBM68}. For calculations of the two-neutron transfer induced by the (t,p) reaction, the reduced radii and diffuseness were set to 1.25 fm and 0.65 fm, respectively, for proton and 1.26 fm and 0.70 fm for deuteron.

In the following, this theoretical approach is applied to the (t,p) and ($^{18}$O,$^{16}$O) reactions.

\begin{table} [H]
\caption{Measured (\textbf{E}$_{exp}$) and shell model calculations (\textbf{E}$_{model}$) energies for low-lying states of $^{28,29,30}$Si nuclei using the \textit{psdmod} interaction.(\textbf{$\Delta$ E}) is defined as \textbf{E}$_{exp}$ - \textbf{E}$_{model}$. } 
\centering
\begin{tabular}{|c|c|c|c|c|} \hline
 \textbf{Nucleus} & J$^{\pi}$&\textbf{E$_{exp}$} & \textbf{E$_{model}$} & $\mid$ $\Delta$ E $\mid$\\
                  &          &     (MeV)         & (MeV)                   & (MeV) \\ \hline
                       
          & 0$^{+}_{1}$ & 0      & 0  & 0\\ \cline{2-5}
$^{28}$Si & 2$^{+}_{1}$ & 1.779  & 2.116 & 0.337 \\ \cline{2-5}        
          & 4$^{+}_{1}$ & 4.617  & 4.753 & 0.136 \\ \hline

          & 1/2$^{+}_{1}$   & 0        & 0 & 0 \\ \cline{2-5}
          & 3/2$^{+}_{1}$ & 1.273  & 0.983 & 0.290\\ \cline{2-5}        
$^{29}$Si & 5/2$^{+}_{1}$ & 2.028  & 2.311 & 0.283\\ \cline{2-5}
          & 3/2$^{+}_{2}$ & 2.425  & 2.614 & 0.189 \\ \cline{2-5}        
          & 5/2$^{+}_{2}$ & 3.067  & 3.332 & 0.265 \\ \hline

          & 0$^{+}_{1}$   & 0    & 0 & 0 \\ \cline{2-5}
          & 2$^{+}_{1}$ & 2.235  & 2.142 & 0.093 \\ \cline{2-5}        
$^{30}$Si & 2$^{+}_{2}$ & 3.498  & 3.512 & 0.014 \\ \cline{2-5}
          & 1$^{+}_{1}$ & 3.769  & 3.775 & 0.006\\ \cline{2-5}        
          & 0$^{+}_{2}$ & 3.788  & 3.529  &0.259\\ \hline
\end{tabular}
\label{energia}
\end{table} 

\begin{table} [H]
\caption{Two-neutron spectroscopic amplitudes for CRC calculations from shell model calculations with \textit{psdmod} \cite{UC11} interaction. n$_1$l$_1$j$_1$ n$_2$l$_2$j$_2$ are the principal quantum numbers, the orbital and the total angular momenta of the neutron 1 and 2 respect to the core. J$_{12}$ is the angular momentum of two-neutron system.} 
\footnotesize
\centering
\begin{tabular}{|c|c|c|c|c|}
\hline \textbf{Initial State} & \textbf{ n$_1$l$_1$j$_1$} & \textbf{Final State} & \textbf{J$_{12}$} & \textbf{Spect. Ampl.}\\ 
                           &  \textbf{ n$_2$l$_2$j$_2$} &                     &            & \\ \hline

                       & $(1p_{3/2})^2$  &                        &   &  0.007\\ \cline{2-2}\cline{5-5} 
                       & $(1p_{1/2})^2$  &                        &   &  0.011\\ \cline{2-2}\cline{5-5}
$^{28}$Si$_{g.s}(0^+)$ & $(1d_{5/2})^2$  & $^{30}$Si$_{g.s}(0^+)$ & 0 &  0.410\\  \cline{2-2}\cline{5-5} 
                       & $(1d_{3/2})^2$  &                        &   &  0.576\\  \cline{2-2}\cline{5-5}
                       & $(2s_{1/2})^2$  &                        &   &  0.518\\ \hline

                       & $(1p_{3/2})^2$          &                        &   & -0.0003\\ \cline{2-2}\cline{5-5} 
                       & $(1p_{3/2})(1p_{1/2})$  &                        &   & -0.007\\ \cline{2-2}\cline{5-5}
$^{28}$Si$_{g.s}(0^+)$ & $(1d_{5/2})^2$      & $^{30}$Si$^*_{2.235}(2^+)$ & 2 & -0.076\\  \cline{2-2}\cline{5-5} 
                       & $(1d_{5/2})(1d_{3/2})$  &                        &   &  0.026\\  \cline{2-2}\cline{5-5}
                       & $(1d_{5/2})(2s_{1/2})$  &                        &   &  0.011\\  \cline{2-2}\cline{5-5}
                       & $(1d_{3/2})^2$          &                        &   & -0.248\\  \cline{2-2}\cline{5-5}
                       & $(1d_{3/2})(2s_{1/2})$  &                        &   & -0.537\\ \hline

                       & $(1p_{3/2})^2$          &                        &   &  0.004\\ \cline{2-2}\cline{5-5} 
                       & $(1p_{3/2})(1p_{1/2})$  &                        &   & -0.005\\ \cline{2-2}\cline{5-5}
$^{28}$Si$_{g.s}(0^+)$ & $(1d_{5/2})^2$      & $^{30}$Si$^*_{3.498}(2^+)$ & 2 &  0.148\\  \cline{2-2}\cline{5-5} 
                       & $(1d_{5/2})(1d_{3/2})$  &                        &   &  0.013\\  \cline{2-2}\cline{5-5}
                       & $(1d_{5/2})(2s_{1/2})$  &                        &   &  0.092\\  \cline{2-2}\cline{5-5}
                       & $(1d_{3/2})^2$          &                        &   &  0.232\\  \cline{2-2}\cline{5-5}
                       & $(1d_{3/2})(2s_{1/2})$  &                        &   &  0.167\\ \hline

                       & $(1p_{3/2})(1p_{1/2})$  &                        &   & 0.001\\ \cline{2-2}\cline{5-5}
$^{28}$Si$_{g.s}(0^+)$ &$(1d_{5/2})(1d_{3/2})$   & $^{30}$Si$^*_{3.769}(1^+)$ & 1 & -0.027\\  \cline{2-2}\cline{5-5} 
                       & $(1d_{3/2})(2s_{1/2})$  &                        &   &  -0.594\\  \hline

                       & $(1p_{3/2})^2$  &                        &   &  -0.003\\ \cline{2-2}\cline{5-5} 
                       & $(1p_{1/2})^2$  &                        &   &  0.013\\ \cline{2-2}\cline{5-5}
$^{28}$Si$_{g.s}(0^+)$ & $(1d_{5/2})^2$  & $^{30}$Si$^*_{3.788}(0^+)$ & 0 &  0.041\\  \cline{2-2}\cline{5-5} 
                       & $(1d_{3/2})^2$  &                        &   &  0.528\\  \cline{2-2}\cline{5-5}
                       & $(2s_{1/2})^2$  &                        &   &  -0.427\\ \hline

                       & $(1p_{3/2})^2$          &                        &   & -0.015\\ \cline{2-2}\cline{5-5} 
                       & $(1p_{3/2})(1p_{1/2})$  &                        &   &  0.037\\ \cline{2-2}\cline{5-5}
$^{28}$Si$^*_{1.779}(2^+)$ & $(1d_{5/2})^2$      & $^{30}$Si$_{gs}(0^+)$  & 2 & -0.325\\  \cline{2-2}\cline{5-5} 
                       & $(1d_{5/2})(1d_{3/2})$  &                        &   &  0.175\\  \cline{2-2}\cline{5-5}
                       & $(1d_{5/2})(2s_{1/2})$  &                        &   & -0.662\\  \cline{2-2}\cline{5-5}
                       & $(1d_{3/2})^2$          &                        &   & -0.095\\  \cline{2-2}\cline{5-5}
                       & $(1d_{3/2})(2s_{1/2})$  &                        &   &  0.073\\ \hline

                       & $(1p_{3/2})^2$  &                            &   &  -0.006\\ \cline{2-2}\cline{5-5} 
                       & $(1p_{1/2})^2$  &                            &   &  -0.010\\ \cline{2-2}\cline{5-5}
$^{28}$Si$^*_{1.779}(2^+)$ & $(1d_{5/2})^2$  & $^{30}$Si$^*_{2.235}(2^+)$ & 0 &  -0.274\\  \cline{2-2}\cline{5-5} 
                       & $(1d_{3/2})^2$  &                            &   &  -0.329\\  \cline{2-2}\cline{5-5}
                       & $(2s_{1/2})^2$  &                            &   &  -0.158\\ \cline{2-2}\cline{4-5}

                       & $(1p_{3/2})(1p_{1/2})$  &                        &   & -0.001\\  \cline{2-2}\cline{5-5}
                       & $(1d_{5/2})(1d_{3/2})$  &                        & 1 & -0.052\\  \cline{2-2}\cline{5-5}
                       & $(1d_{3/2})(2s_{1/2})$  &                        &   &  0.012\\ \cline{2-2}\cline{4-5}

                       & $(1p_{3/2})^2$          &                        &   & -0.004\\ \cline{2-2}\cline{5-5} 
                       & $(1p_{3/2})(1p_{1/2})$  &                        &   &  0.006\\ \cline{2-2}\cline{5-5}
                       & $(1d_{5/2})^2$          &                        & 2 & -0.212\\  \cline{2-2}\cline{5-5} 
                       & $(1d_{5/2})(1d_{3/2})$  &                        &   & -0.042\\  \cline{2-2}\cline{5-5}
                       & $(1d_{5/2})(2s_{1/2})$  &                        &   &  0.014\\  \cline{2-2}\cline{5-5}
                       & $(1d_{3/2})^2$          &                        &   & -0.180\\  \cline{2-2}\cline{5-5}
                       & $(1d_{3/2})(2s_{1/2})$  &                        &   &  0.032\\ \cline{2-2}\cline{4-5}
 
                       & $(1d_{5/2})(1d_{3/2})$  &                        & 3 & -0.229\\  \cline{2-2}\cline{5-5}
                       & $(1d_{5/2})(2s_{1/2})$  &                        &   &  0.143\\  \cline{2-2}\cline{4-5}

                       & $(1d_{5/2})^2$          &                        & 4 &  0.088\\  \cline{2-2}\cline{5-5} 
                       & $(1d_{5/2})(1d_{3/2})$  &                        &   & -0.624\\  \hline

\end{tabular}
\label{table-ic}
\end{table}

\begin{table} [H]
\caption{Continuation of Table \ref{table-ic}} 
\footnotesize
\centering
\begin{tabular}{|c|c|c|c|c|}
\hline \textbf{Initial State} & \textbf{ n$_1$l$_1$j$_1$} & \textbf{Final State} & \textbf{J$_{12}$} & \textbf{Spect. Ampl.}\\ 
                           &  \textbf{ n$_2$l$_2$j$_2$} &                     &            & \\ \hline

                       & $(1p_{3/2})^2$  &                            &   &  -0.007\\ \cline{2-2}\cline{5-5} 
                       & $(1p_{1/2})^2$  &                            &   &  -0.014\\ \cline{2-2}\cline{5-5}
$^{28}$Si$^*_{1.779}(2^+)$ & $(1d_{5/2})^2$  & $^{30}$Si$^*_{3.498}(2^+)$ & 0 &  -0.275\\  \cline{2-2}\cline{5-5} 
                       & $(1d_{3/2})^2$  &                            &   &  -0.376\\  \cline{2-2}\cline{5-5}
                       & $(2s_{1/2})^2$  &                            &   &  -0.283\\ \cline{2-2}\cline{4-5}

                       & $(1p_{3/2})(1p_{1/2})$  &                        &   & -0.00003\\  \cline{2-2}\cline{5-5}
                       & $(1d_{5/2})(1d_{3/2})$  &                        & 1 & -0.044\\  \cline{2-2}\cline{5-5}
                       & $(1d_{3/2})(2s_{1/2})$  &                        &   &  0.112\\ \cline{2-2}\cline{4-5}

                       & $(1p_{3/2})^2$          &                        &   & -0.003\\ \cline{2-2}\cline{5-5} 
                       & $(1p_{3/2})(1p_{1/2})$  &                        &   &  0.015\\ \cline{2-2}\cline{5-5}
                       & $(1d_{5/2})^2$          &                        & 2 &  0.045\\  \cline{2-2}\cline{5-5} 
                       & $(1d_{5/2})(1d_{3/2})$  &                        &   &  0.003\\  \cline{2-2}\cline{5-5}
                       & $(1d_{5/2})(2s_{1/2})$  &                        &   & -0.213\\  \cline{2-2}\cline{5-5}
                       & $(1d_{3/2})^2$          &                        &   &  0.209\\  \cline{2-2}\cline{5-5}
                       & $(1d_{3/2})(2s_{1/2})$  &                        &   &  0.242\\ \cline{2-2}\cline{4-5}
 
                       & $(1d_{5/2})(1d_{3/2})$  &                        & 3 &  0.188\\  \cline{2-2}\cline{5-5}
                       & $(1d_{5/2})(2s_{1/2})$  &                        &   & -0.159\\  \cline{2-2}\cline{4-5}

                       & $(1d_{5/2})^2$          &                        & 4 & -0.132\\  \cline{2-2}\cline{5-5} 
                       & $(1d_{5/2})(1d_{3/2})$  &                        &   &  0.093\\  \hline

                       & $(1p_{3/2})(1p_{1/2})$  &                        &   &  0.003\\  \cline{2-2}\cline{5-5}
                       & $(1d_{5/2})(1d_{3/2})$  &                        & 1 &  0.229\\  \cline{2-2}\cline{5-5}
                       & $(1d_{3/2})(2s_{1/2})$  &                        &   & -0.094\\ \cline{2-2}\cline{4-5}
                       
                       & $(1p_{3/2})^2$          &                        &   &  0.001\\ \cline{2-2}\cline{5-5} 
                       & $(1p_{3/2})(1p_{1/2})$  &                        &   & -0.001\\ \cline{2-2}\cline{5-5}
$^{28}$Si$^*_{1.779}(2^+)$ & $(1d_{5/2})^2$      & $^{30}$Si$^*_{3.769}(1^+)$& 2 &  0.139\\  \cline{2-2}\cline{5-5} 
                       & $(1d_{5/2})(1d_{3/2})$  &                        &   &  0.396\\  \cline{2-2}\cline{5-5}
                       & $(1d_{5/2})(2s_{1/2})$  &                        &   &  0.256\\  \cline{2-2}\cline{5-5}
                       & $(1d_{3/2})^2$          &                        &   & -0.296\\  \cline{2-2}\cline{5-5}
                       & $(1d_{3/2})(2s_{1/2})$  &                        &   &  0.115\\  \cline{2-2}\cline{4-5}

                       & $(1d_{5/2})(1d_{3/2})$  &                        & 3 &  0.483\\  \cline{2-2}\cline{5-5}
                       & $(1d_{5/2})(2s_{1/2})$  &                        &   & -0.134\\  \hline

                       & $(1p_{3/2})^2$          &                        &   & -0.002\\ \cline{2-2}\cline{5-5} 
                       & $(1p_{3/2})(1p_{1/2})$  &                        &   &  0.009\\ \cline{2-2}\cline{5-5}
$^{28}$Si$^*_{1.779}(2^+)$ & $(1d_{5/2})^2$      & $^{30}$Si$^*_{3.788}(0^+)$  & 2 & 0.162\\  \cline{2-2}\cline{5-5} 
                       & $(1d_{5/2})(1d_{3/2})$  &                        &   &  0.406\\  \cline{2-2}\cline{5-5}
                       & $(1d_{5/2})(2s_{1/2})$  &                        &   &  0.586\\  \cline{2-2}\cline{5-5}
                       & $(1d_{3/2})^2$          &                        &   &  0.190\\  \cline{2-2}\cline{5-5}
                       & $(1d_{3/2})(2s_{1/2})$  &                        &   & -0.112\\ \hline

                       & $(1d_{5/2})^2$  &                                   &   &   0.025\\ \cline{2-2}\cline{5-5}
$^{28}$Si$^*_{4.617}(4^+)$ & $(1d_{5/2})(1d_{3/2})$  & $^{30}$Si$_{gs}(0^+)$ & 4 &  -0.887\\  \hline
                       
                       & $(1p_{3/2})^2$          &                        &   & -0.011\\ \cline{2-2}\cline{5-5} 
                       & $(1p_{3/2})(1p_{1/2})$  &                        &   & -0.029\\ \cline{2-2}\cline{5-5}
$^{28}$Si$^*_{4.617}(4^+)$ & $(1d_{5/2})^2$      & $^{30}$Si$^*_{2.235}(2^+)$ & 2 & -0.196\\  \cline{2-2}\cline{5-5} 
                       & $(1d_{5/2})(1d_{3/2})$  &                        &   &  0.184\\  \cline{2-2}\cline{5-5}
                       & $(1d_{5/2})(2s_{1/2})$  &                        &   & -0.609\\  \cline{2-2}\cline{5-5}
                       & $(1d_{3/2})^2$          &                        &   & -0.119\\  \cline{2-2}\cline{5-5}
                       & $(1d_{3/2})(2s_{1/2})$  &                        &   & -0.157\\ \cline{2-2} \cline{4-5}

                       & $(1d_{5/2})(1d_{3/2})$  &                        & 3 &  0.186\\  \cline{2-2}\cline{5-5}
                       & $(1d_{5/2})(2s_{1/2})$  &                        &   & -0.306\\  \cline{2-2}\cline{4-5}

                       & $(1d_{5/2})^2$          &                        & 4 & -0.118\\  \cline{2-2}\cline{5-5} 
                       & $(1d_{5/2})(1d_{3/2})$  &                        &   &  0.210\\  \hline

\end{tabular}
\label{table-ic2}
\end{table}

\begin{table} [H]
\caption{Continuation of Table \ref{table-ic2}} .
\footnotesize
\centering
\begin{tabular}{|c|c|c|c|c|}
\hline \textbf{Initial State} & \textbf{ n$_1$l$_1$j$_1$} & \textbf{Final State} & \textbf{J$_{12}$} & \textbf{Spect. Ampl.}\\ 
                           &  \textbf{ n$_2$l$_2$j$_2$} &                     &            & \\ \hline

                       & $(1p_{3/2})^2$          &                        &   & -0.005\\ \cline{2-2}\cline{5-5} 
                       & $(1p_{3/2})(1p_{1/2})$  &                        &   &  0.009\\ \cline{2-2}\cline{5-5}
$^{28}$Si$^*_{4.617}(4^+)$ & $(1d_{5/2})^2$      & $^{30}$Si$^*_{3.498}(2^+)$ & 2 & -0.086\\  \cline{2-2}\cline{5-5} 
                       & $(1d_{5/2})(1d_{3/2})$  &                        &   &  -0.033\\  \cline{2-2}\cline{5-5}
                       & $(1d_{5/2})(2s_{1/2})$  &                        &   & -0.140\\  \cline{2-2}\cline{5-5}
                       & $(1d_{3/2})^2$          &                        &   & -0.089\\  \cline{2-2}\cline{5-5}
                       & $(1d_{3/2})(2s_{1/2})$  &                        &   &  0.014\\ \cline{2-2} \cline{4-5}
                       
                       & $(1d_{5/2})(1d_{3/2})$  &                        & 3 & -0.198\\  \cline{2-2}\cline{5-5}
                       & $(1d_{5/2})(2s_{1/2})$  &                        &   &  0.059\\  \cline{2-2}\cline{4-5}

                       & $(1d_{5/2})^2$          &                        & 4 &  0.031\\  \cline{2-2}\cline{5-5} 
                       & $(1d_{5/2})(1d_{3/2})$  &                        &   & -0.518\\  \hline

                       & $(1d_{5/2})(1d_{3/2})$  &                        & 3 & -0.176\\  \cline{2-2}\cline{5-5}
$^{28}$Si$^*_{4.617}(4^+)$&$(1d_{5/2})(2s_{1/2})$& $^{30}$Si$^*_{3.769}(1^+)$&   &  0.575\\  \cline{2-2}\cline{4-5} 
                       
                       & $(1d_{5/2})^2$  &                                   &   &   0.237\\ \cline{2-2}\cline{5-5}
                       & $(1d_{5/2})(1d_{3/2})$  &                        & 4 &  -0.065\\  \hline
 
                       & $(1d_{5/2})^2$  &                                   &   &   0.335\\ \cline{2-2}\cline{5-5}
$^{28}$Si$^*_{4.617}(4^+)$ & $(1d_{5/2})(1d_{3/2})$  & $^{30}$Si$^*_{3.788}(0^+)$ & 4 &0.627\\  \hline
                      
\end{tabular}
\label{table-ic3}
\end{table}

\begin{table} [H]
\caption{One-neutron spectroscopic amplitudes, adopted in CCBA cross section calculations for $^{28}$Si to $^{29}$Si transitions, obtained by shell model with the \textit{psdmod} \cite{UC11} interaction. nl$_j$ are the principal quantum number, the orbital and the total angular momentum of the single neutron, respectively.} 
\centering
\begin{tabular}{|c|c|c|c|}
\hline \textbf{Initial State} & \textbf{nl$_j$} & \textbf{Final State} & \textbf{Spect. Ampl.}\\ \hline

                         & $(2s_{1/2})$  & $^{29}$Si$_{gs}(1/2^+_1)$      &  0.716\\ \cline{2-4}
$^{28}$Si$_{g.s}(0^+_1)$   & $(1d_{3/2})$  & $^{29}$Si$^*_{1.273}(3/2^+_1)$ & -0.828\\ \cline{2-4}
                         & $(1d_{5/2})$  & $^{29}$Si$^*_{2.028}(5/2^+_1)$ & -0.347\\  \cline{2-4}
                         & $(1d_{3/2})$  & $^{29}$Si$^*_{2.425}(3/2^+_2)$ &  0.046\\  \cline{2-4}
                         & $(1d_{5/2})$  & $^{29}$Si$^*_{3.067}(5/2^+_2)$ & -0.226\\ \hline

                         & $(1d_{3/2})$  & $^{29}$Si$_{gs}(1/2^+_1)$      & -0.388\\ \cline{2-2}\cline{4-4} 
                         & $(1d_{5/2})$  &                              & -0.847\\ \cline{2-4}
$^{28}$Si$^*_{1.779}(2^+_1)$ & $(2s_{1/2})$&                              & -0.089\\ \cline{2-2} \cline{4-4}
                         & $(1d_{3/2})$  & $^{29}$Si$^*_{1.273}(3/2^+_1)$ & -0.006\\  \cline{2-2}\cline{4-4} 
                         & $(1d_{5/2})$  &                              &  0.293\\  \cline{2-4}
                           & $(2s_{1/2})$&                              &  0.631\\ \cline{2-2} \cline{4-4}
                         & $(1d_{3/2})$  & $^{29}$Si$^*_{2.028}(5/2^+_1)$ &  0.025\\  \cline{2-2}\cline{4-4} 
                         & $(1d_{5/2})$  &                              &  0.414\\  \cline{2-4}
                         & $(2s_{1/2})$&                                & -0.342\\ \cline{2-2} \cline{4-4}
                         & $(1d_{3/2})$  & $^{29}$Si$^*_{2.425}(3/2^+_2)$ &  0.748\\  \cline{2-2}\cline{4-4} 
                         & $(1d_{5/2})$  &                              & -0.518\\  \cline{2-4}
                         & $(2s_{1/2})$&                                &  0.013\\ \cline{2-2} \cline{4-4}
                         & $(1d_{3/2})$  & $^{29}$Si$^*_{3.067}(5/2^+_2)$ & -0.761\\  \cline{2-2}\cline{4-4} 
                         & $(1d_{5/2})$  &                              &  0.051\\  \hline

                         & $(1d_{5/2})$  & $^{29}$Si$^*_{1.273}(3/2^+_1)$ &  0.904\\  \cline{2-4}                            
                         & $(1d_{3/2})$  & $^{29}$Si$^*_{2.028}(5/2^+_1)$ & -0.425\\  \cline{2-2}\cline{4-4} 
$^{28}$Si$^*_{4.617}(4^+_1)$&$(1d_{5/2})$  &                              & -0.195\\  \cline{2-4}
                         & $(1d_{5/2})$  & $^{29}$Si$^*_{2.425}(3/2^+_2)$ &  0.295\\  \cline{2-4}
                         & $(1d_{3/2})$  & $^{29}$Si$^*_{3.067}(5/2^+_2)$ & -0.244\\  \cline{2-2}\cline{4-4} 
                         & $(1d_{5/2})$  &                              &  0.406\\  \hline

\end{tabular}
\label{seq}
\end{table}

\begin{table} [H]
\caption{One-neutron spectroscopic amplitudes, adopted in CCBA cross section calculations for $^{28}$Si to $^{29}$Si transitions, obtained by shell model with the \textit{psdmod} \cite{UC11} interaction. nl$_j$ are the principal quantum number, the orbital and the total angular momentum of the single neutron, respectively.} 
\centering
\begin{tabular}{|c|c|c|c|}
\hline \textbf{Initial State} & \textbf{nl$_j$} & \textbf{Final State} & \textbf{Spect. Ampl.}\\ \hline                  
                  
                         & $(2s_{1/2})$  & $^{30}$Si$_{gs}(0^+_1)$      & -0.867\\ \cline{2-4}
$^{29}$Si$_{g.s}(1/2^+_1)$ & $(1d_{3/2})$  & $^{30}$Si$^*_{2.235}(2^+_1)$ &  0.768\\ \cline{2-2} \cline{4-4}
                         & $(1d_{5/2})$  &                            & -0.232\\  \cline{2-4}
                         & $(1d_{3/2})$  & $^{30}$Si$^*_{3.498}(2^+_2)$ & -0.205\\  \cline{2-2} \cline{4-4}
                         & $(1d_{5/2})$  &                            & -0.069\\ \cline{2-4}
                         & $(2s_{1/2})$  & $^{30}$Si$^*_{3.769}(1^+_1)$ & -0.001\\  \cline{2-2} \cline{4-4}
                         & $(1d_{3/2})$  &                            & -0.792\\ \cline{2-4}
                         & $(2s_{1/2})$  & $^{30}$Si$^*_{3.788}(0^+_2)$   & -0.731\\ \hline

                         & $(1d_{3/2})$  & $^{30}$Si$_{gs}(0^+_1)$      &  0.973\\ \cline{2-4}
                         & $(2s_{1/2})$  &                            &  0.703\\ \cline{2-2}\cline{4-4}
$^{29}$Si$^*_{1.273}(3/2^+_1)$& $(1d_{3/2})$& $^{30}$Si$^*_{2.235}(2^+_1)$& -0.438\\ \cline{2-2} \cline{4-4}
                         & $(1d_{5/2})$  &                            & -0.013\\  \cline{2-4}
                         & $(2s_{1/2})$  &                            & -0.106\\ \cline{2-2}\cline{4-4} 
                         & $(1d_{3/2})$  & $^{30}$Si$^*_{3.498}(2^+_2)$ &  0.423\\  \cline{2-2} \cline{4-4}
                         & $(1d_{5/2})$  &                            &  0.031\\ \cline{2-4}
                         & $(2s_{1/2})$  & $^{30}$Si$^*_{3.769}(1^+_1)$ &  0.734\\  \cline{2-2} \cline{4-4}
                         & $(1d_{3/2})$  &                            & -0.026\\  \cline{2-2} \cline{4-4}
                         & $(1d_{5/2})$  &                            &  0.090\\ \cline{2-4}
                         & $(1d_{3/2})$  & $^{30}$Si$^*_{3.788}(0^+_2)$   &  0.851\\ \hline

                         & $(1d_{5/2})$  & $^{30}$Si$_{gs}(0^+_1)$      &  1.370\\ \cline{2-4}
                         & $(2s_{1/2})$  &                            &  0.124\\ \cline{2-2}\cline{4-4}
$^{29}$Si$^*_{2.028}(5/2^+_1)$& $(1d_{3/2})$& $^{30}$Si$^*_{2.235}(2^+_1)$& -0.086\\ \cline{2-2} \cline{4-4}
                         & $(1d_{5/2})$  &                            & -0.025\\  \cline{2-4}
                         & $(2s_{1/2})$  &                            &  0.311\\ \cline{2-2}\cline{4-4} 
                         & $(1d_{3/2})$  & $^{30}$Si$^*_{3.498}(2^+_2)$ & -0.436\\  \cline{2-2} \cline{4-4}
                         & $(1d_{5/2})$  &                            &  0.446\\ \cline{2-4}
                         & $(1d_{3/2})$  & $^{30}$Si$^*_{3.769}(1^+_1)$ & -0.108\\  \cline{2-2} \cline{4-4}
                         & $(1d_{5/2})$  &                            &  0.311\\ \cline{2-4}
                         & $(1d_{5/2})$  & $^{30}$Si$^*_{3.788}(0^+_2)$   & -0.532\\ \hline

                         & $(1d_{3/2})$  & $^{30}$Si$_{gs}(0^+_1)$      &  0.145\\ \cline{2-4}
                         & $(2s_{1/2})$  &                            &  0.091\\ \cline{2-2}\cline{4-4}
$^{29}$Si$^*_{2.425}(3/2^+_2)$& $(1d_{3/2})$& $^{30}$Si$^*_{2.235}(2^+_1)$&  0.242\\ \cline{2-2} \cline{4-4}
                         & $(1d_{5/2})$  &                            &  0.528\\  \cline{2-4}
                         & $(2s_{1/2})$  &                            &  0.490\\ \cline{2-2}\cline{4-4} 
                         & $(1d_{3/2})$  & $^{30}$Si$^*_{3.498}(2^+_2)$ &  0.316\\  \cline{2-2} \cline{4-4}
                         & $(1d_{5/2})$  &                            & -0.091\\ \cline{2-4}
                         & $(2s_{1/2})$  & $^{30}$Si$^*_{3.769}(1^+_1)$ & -0.216\\  \cline{2-2} \cline{4-4}
                         & $(1d_{3/2})$  &                            & -0.371\\  \cline{2-2} \cline{4-4}
                         & $(1d_{5/2})$  &                            &  0.392\\ \cline{2-4}
                         & $(1d_{3/2})$  & $^{30}$Si$^*_{3.788}(0^+_2)$ & -0.437\\ \hline

                         & $(1d_{5/2})$  & $^{30}$Si$_{gs}(0^+_1)$      &  0.401\\ \cline{2-4}
                         & $(2s_{1/2})$  &                            & -0.079\\ \cline{2-2}\cline{4-4}
$^{29}$Si$^*_{2.028}(5/2^+_2)$& $(1d_{3/2})$& $^{30}$Si$^*_{2.235}(2^+_1)$&  0.071\\ \cline{2-2} \cline{4-4}
                         & $(1d_{5/2})$  &                            & -0.320\\  \cline{2-4}
                         & $(2s_{1/2})$  &                            & -0.137\\ \cline{2-2}\cline{4-4} 
                         & $(1d_{3/2})$  & $^{30}$Si$^*_{3.498}(2^+_2)$ &  0.630\\  \cline{2-2} \cline{4-4}
                         & $(1d_{5/2})$  &                            &  0.226\\ \cline{2-4}
                         & $(1d_{3/2})$  & $^{30}$Si$^*_{3.769}(1^+_1)$ &  0.311\\  \cline{2-2} \cline{4-4}
                         & $(1d_{5/2})$  &                            & -0.550\\ \cline{2-4}
                         & $(1d_{5/2})$  & $^{30}$Si$^*_{3.788}(0^+_2)$   &  0.724\\ \hline                      

\end{tabular}
\label{seq1}
\end{table}

%
%
\subsubsection*{$^{28}$Si(t,p)$^{30}$Si at 18 MeV}

The experimental angular distributions for the (t,p) reaction at 18 MeV were obtained from Ref. \cite{ACH86}. 

\begin{figure}[ht!]
\centering
\graphicspath{{figuras/}}
\includegraphics[width=0.45\textwidth]{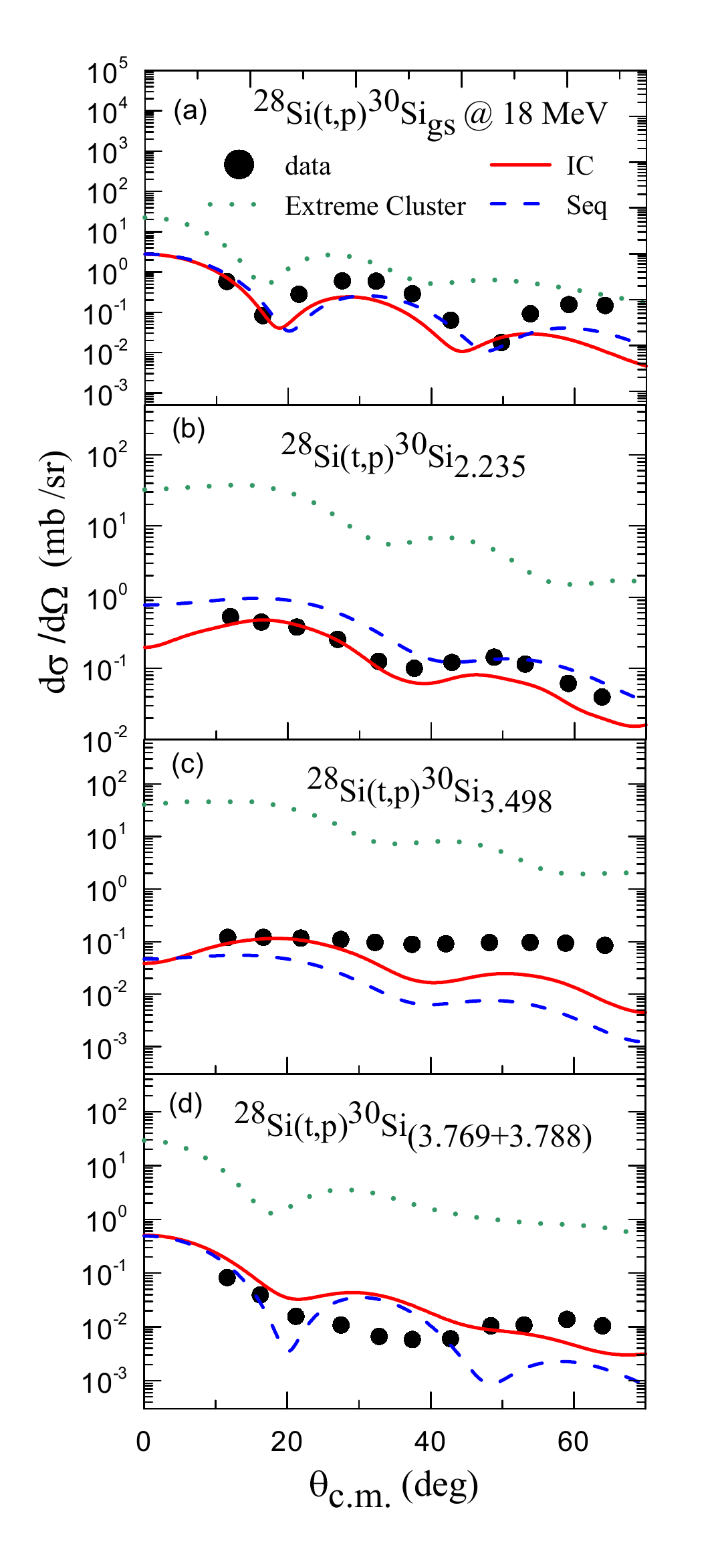}
\caption{(color online) Comparison of the angular distribution of $^{28}$Si(t,p)$^{30}$Si reaction at 18 MeV \cite{ACH86} for the transition to the ground state (a), first excited state at 2.235 MeV (b), second excited state at 3.398 MeV (c) and the sum of the 3.769 MeV and the 3.788 MeV states (d) in $^{30}$Si nuclei.  The green dotted, red solid and blue dashed curves correspond to the extreme cluster, independent coordinates and sequential model, respectively.}
\label{tp-18mev}
\end{figure}

The comparison of the theoretical curves with experimental data are shown in Fig.~\ref{tp-18mev}. The extreme cluster model gives absolute cross sections that are much higher than the other models presented, except for the transition to the ground state of $^{30}$Si.  The oscillatory behavior is reasonably well reproduced by the IC and the sequential model.  This is a first indication that direct and sequential are competing transfer mechanisms feeding $^{30}$Si states.

In Ref. \cite{BaH73} cross section angular distributions of the $^{28}$Si(t,p) reactions at 10.5 and 12 MeV are reported in arbitrary units. We performed calculations at these energies and obtained a good reproduction of the experimental shapes.

The calculated angular distributions, shown in Fig.~\ref{tp-18mev}  for transitions to low-lying states of $^{30}$Si, nicely reproduce the experimental data. This result, together with the accurate description of the excitation energies for the same states (see Table \ref{energia}), support the choice of the \textit{psdmod} interaction and the valence space in our shell model calculations.
For this reason the same nuclear structure inputs were used in the ($^{18}$O,$^{16}$O), discussed in the next section.

\subsubsection*{$^{28}$Si($^{18}$O,$^{16}$O)$^{30}$Si at 56 MeV}

The experimental angular distributions of two-neutron transfer cross sections induced by $^{18}$O on $^{28}$Si at 56 MeV bombarding energy were reported in Ref. \cite{MFG79}. In that article, DWBA calculations using an optical potential adjusted to the elastic scattering data, were not able to reproduce the absolute cross section for transfer reactions, requiring arbitrary scaling in the calculations.

\begin{figure}[H]
\graphicspath{{figuras/}}
\includegraphics[width=0.45\textwidth]{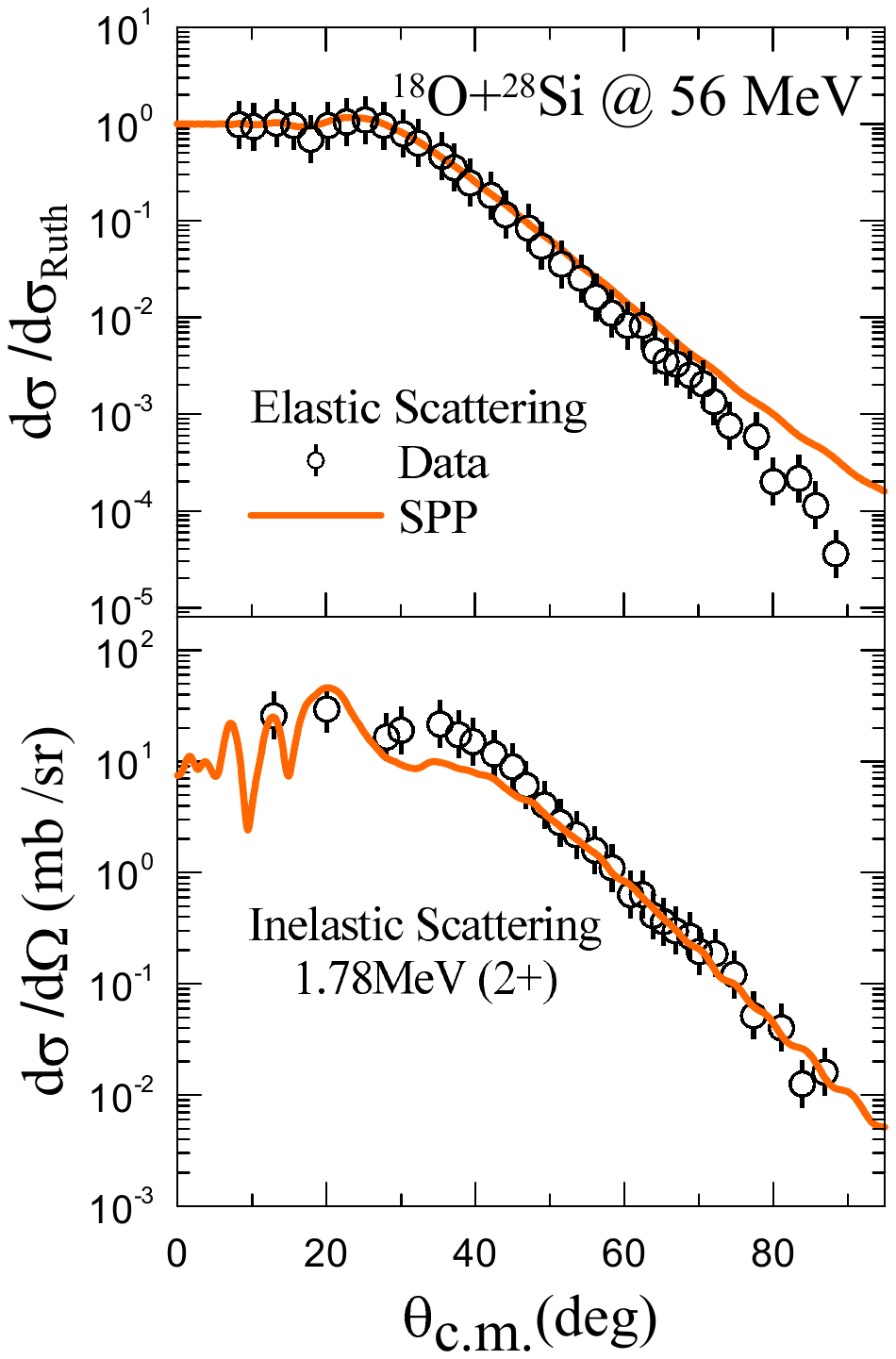}
\caption{(color online) Comparison between experimental and theoretical angular distributions for elastic and inelastic scattering for ($^{18}$O +$^{28}$Si) at 56 MeV \cite{MFG79}. The SSP curve stands for optical potential with the shape of S\~ao Paulo potential with 1.0 and 0.6 factors in the real and imaginary terms.} 
\label{elastic}
\end{figure}

A reanalysis of elastic and inelastic scattering data at 56 MeV, from Ref. \cite{MFG79}, was performed within the theoretical framework discussed in this work in order to assess the quality of the optical potential used in the entrance partition. Comparisons between experimental data and theoretical curves are shown in Fig.~\ref{elastic}. We observe that the 0.6 strength factor to the imaginary part of the optical potential describes reasonably well the elastic and inelastic scattering. This strength factor was varied between  0.2 and 0.6 with minor changes on the calculated transfer cross sections. Therefore, we adopted the 0.6 imaginary factor throughout this work, as done in the previous works \cite{CCB13,ECL16,CFC17,ELL17}.


\begin{figure}[H]
\graphicspath{{figuras/}}
\includegraphics[width=0.45\textwidth]{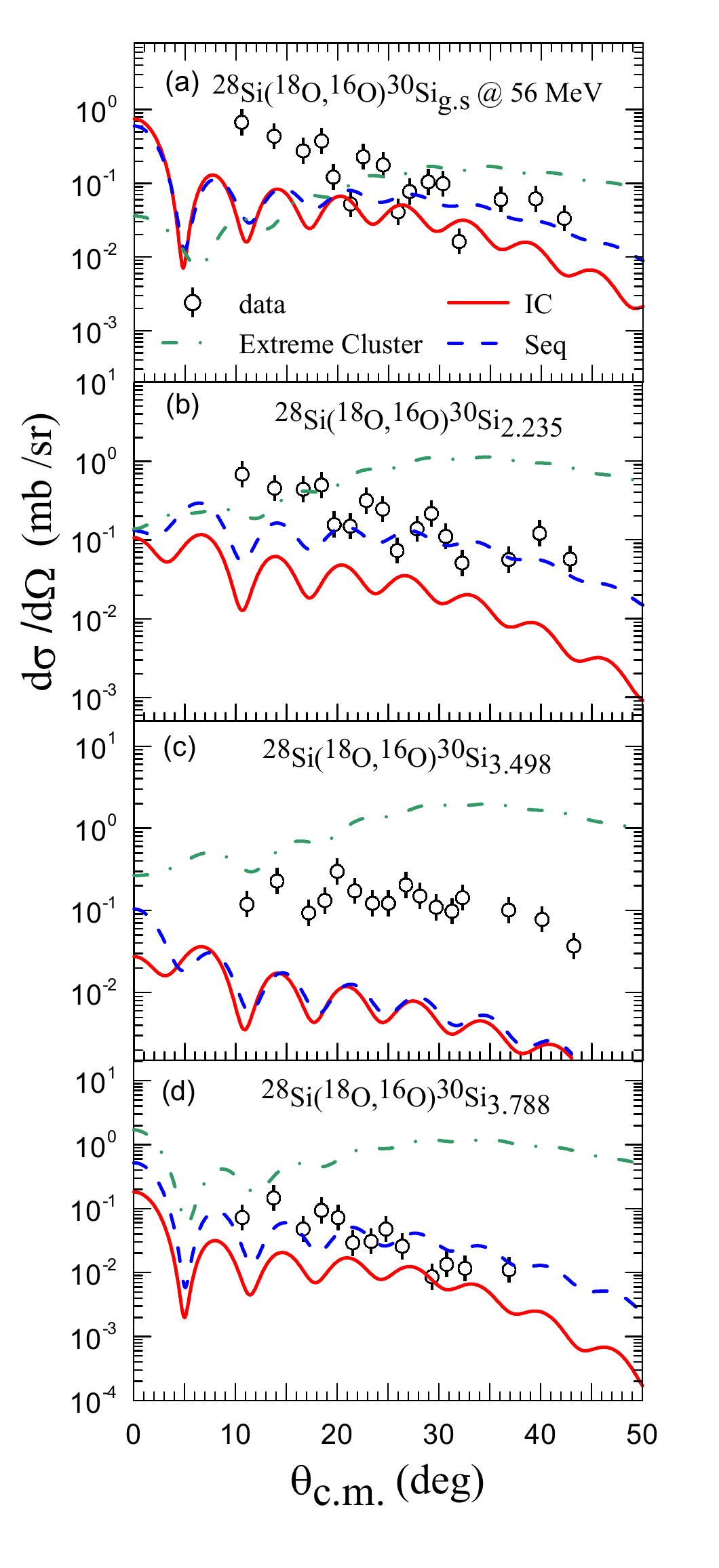}
\caption{(color online) Angular distributions for the transfer of two neutrons to the ground state (a), first excited state 2.235 MeV (b), second excited state 3.398 MeV (c) and fourth excited state 3.788 MeV (d) for the $^{28}$Si($^{18}$O,$^{16}$O)$^{30}$Si reaction  at 56 MeV \cite{MFG79}. The green dotted, red solid and blue dashed curves represent the extreme cluster, the independent coordinates and the sequential models, respectively.}
\label{transf-56MeV}
\end{figure}

The comparison between experimental data and the differential cross sections for the $^{28}$Si($^{18}$O,$^{16}$O)$^{30}$Si reaction at 56 MeV are shown in Fig. \ref{transf-56MeV}. The direct transfer of two neutrons is calculated by CRC approach, using both the extreme cluster and  the independent coordinates models. The sequential transfer cross sections are obtained by second-order CCBA calculation.


At incident energy of 56 MeV the cross section of the two-neutron transfer to the ground state (see Figure \ref{transf-56MeV}a) does not show a good agreement with the experimental data, specially at forward angles, maybe due to the concurrence of the three mechanisms. For the other states, the extreme cluster model overestimates the experimental data. For the states of 2.235 MeV and 3.398 MeV (see Figure \ref{transf-56MeV}b and d) the two-step sequential model describes better the transfer cross section. These results are different to those published in Ref.~\cite{MFG79} where the two-step process was found to be much smaller than the direct process. For the trasnsition to the state at 3.398 MeV (see Figure \ref{transf-56MeV}c) the direct and two-step process are about one order of magnitude below of the data, indicating that a larger base in the model space might be needed to describe it. 
\subsubsection*{$^{28}$Si($^{18}$O,$^{16}$O)$^{30}$Si at 84 MeV}

\begin{figure}[H]
\graphicspath{{figuras/}}
\includegraphics[width=0.45\textwidth]{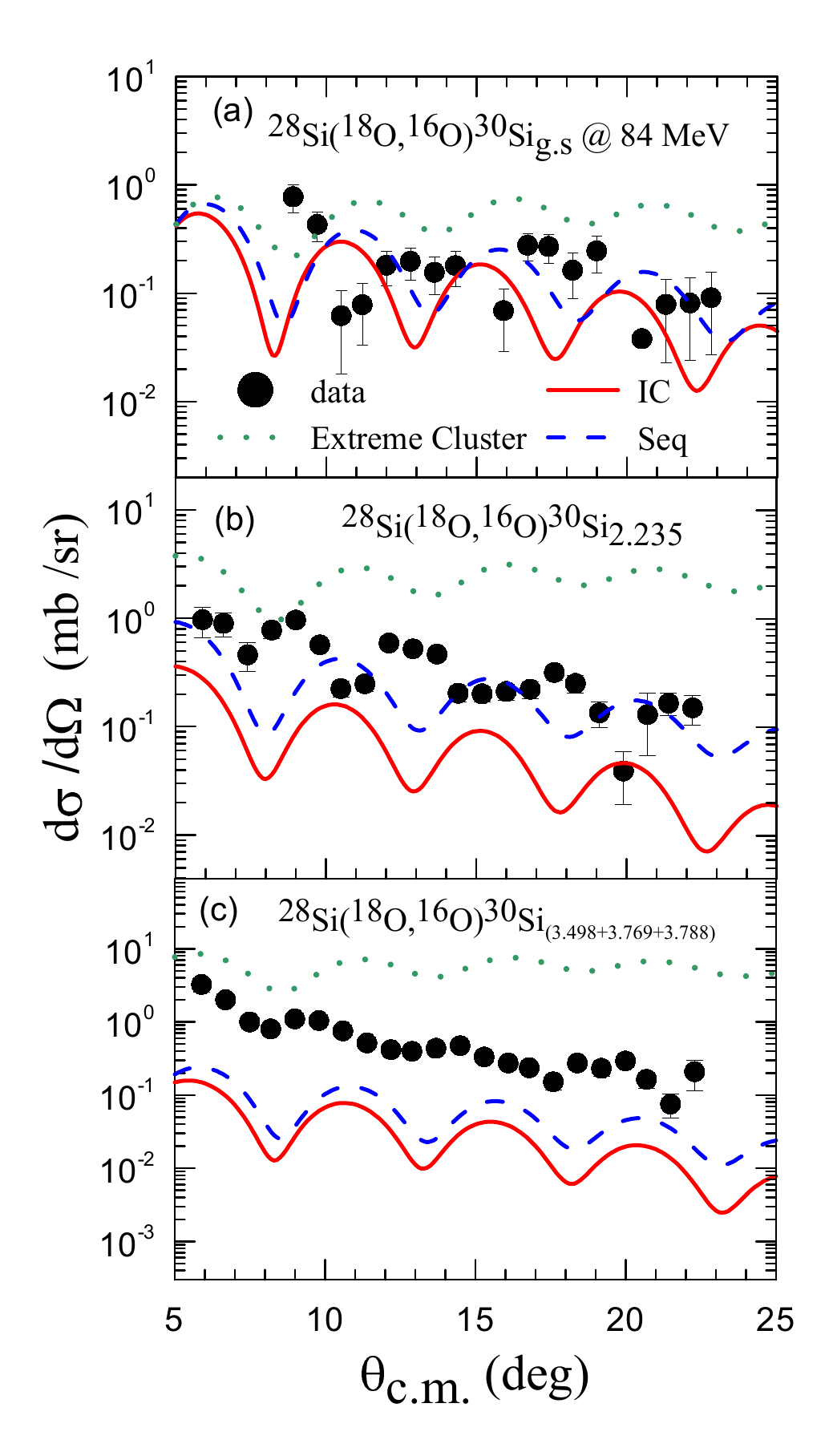}
\caption{Comparison of the experimental angular distribution with the extreme cluster, independent coordinates and sequential models calculations for the $^{28}$Si($^{18}$O,$^{16}$O)$^{30}$Si reaction at 84 MeV.}
\label{ic+seq+cluster}
 \end{figure}

The comparison between the results of the transfer reaction cross sections for the extreme cluster, independent coordinates and sequential models and the experimental data at 84 MeV are shown in Fig. \ref{ic+seq+cluster}. The cross sections shown in Fig. \ref{ic+seq+cluster}c correspond to the sum of the results for the $2^{+}$ (at 3.498 MeV), $1^{+}$ (at 3.769 MeV) and $0^{+}$ (at 3.788 MeV) states which are not resolved in data. The extreme cluster model overestimates all the experimental angular distributions, as previously observed for the data at 56 MeV. A similar behaviour is also observed in the two-neutron transfer to $^{64}$Ni nuclei \cite{PSM17}. This may be related to the oversimplified structure of the two-neutron cluster, with contributions to the wave functions coming only from the component with anti-parallel spin and spectroscopic amplitudes set to 1.0. 

Both the independent coordinates and sequential calculations describe with good accuracy the average absolute cross sections for the ground state of $^{30}$Si nucleus (see Fig. \ref{ic+seq+cluster}a). For the first excited state (2$^{+}$) at 2.235 MeV (see Fig. \ref{ic+seq+cluster}b) the results of sequential model give a better description of the cross section. For the sum of states, 3.498 MeV + 3.769 MeV + 3.788 MeV, both independent coordinates and sequential models results are bellow the experimental data (see Fig. \ref{ic+seq+cluster}c). A possible justification for neither model describing well the magnitude of this cross section is that a larger model space might be necessary. However due to computational limitations, it was not possible to enlarge it.

One can observe in Figure~\ref{ic+seq+cluster} that the theoretical angular distributions are shifted approximately by three degrees in with respect to the experimental data. The fact that the sequential transfer is dominant compared to the simultaneous one can be associated to the axial deformation of $^{28}$Si ground state which emphasize long-range correlations in the many-body wave functions. In these conditions short-range correlations, as those associated to neutron pairing, tend to be less visible. In this way, the response of the nucleus to direct one-step mechanism is weakened due to the selectivity of pairing configurations, and the two-step mechanism is dominant. The results obtained here are in agreement to the conclusions for the transfer to ground and first excited state of $^{26}$Mg \cite{BRP79}, $^{76}$Ge \cite{LeL77} and recently for $^{66}$Ni \cite{PSM17}. They are however different from what found in recent works as Refs. \cite{ECL16,CFC17} for $^{18}$O, $^{14}$C and $^{15}$C, where the direct mechanism is dominant. The long-range correlations are related to the quadrupole deformation, which in the case of $^{30}$Si and $^{66}$Ni is greater than $^{18}$O and $^{14,15}$C, making the effects of collective correlations being evidenced in the former systems. 

 \begin{table}[H]
  \caption{Reduced electric quadrupole transition probabilities for some selected nuclei \cite{RNT01}.}
     \centering

     \begin{tabular}{|c|c|} 
    \hline
         Nucleus & B(E2); \\
                 &0$^+$ $\rightarrow$ 2$^+$  (e$^2$b$^2$)\\ \hline
     
$^{14}$C & 0.0018\\ \hline
$^{18}$O & 0.0045\\ \hline
$^{28}$Mg & 0.035 \\ \hline
$^{30}$Si & 0.022 \\ \hline
$^{66}$Ni & 0.060 \\ \hline
$^{76}$Ge & 0.270\\ \hline

         \hline
     \end{tabular}
     
     \label{beta}
  \end{table}

In Table \ref{beta} we list the B(E2) (reduced electric quadrupole transition probability) between the ground and the first (2+) excited state of nuclei studied by the ($^{18}$O,$^{16}$O) reaction. For cases where the simultaneous two-neutron transfer mechanism is dominant, the B(E2) is small, as in case of $^{18}$O and $^{14}$C. However, when the composite nucleus is deformed, as in the $^{30}$Si case, the short-range pairing correlations between the two neutrons  do not prevail. This means that the two-neutron transfer mechanism is intrinsically related to the nuclear structure of the target nucleus. The effect of different bombarding energies on the competition between simultaneous and sequential mechanisms have been probed in our calculations at energies below and well above the Coulomb barrier leading to similar conclusions. We also performed CRC and two-step CCBA calculations at higher energies (120 MeV) and observed that the conclusions are similar. 


\subsection{Microscopic Cluster Model}

As mentioned in the previous sub-sections, in the extreme cluster model the pair of neutrons are transferred with spin anti-parallel ($S = 0$) and spectroscopic amplitudes set to 1.0. This approximation overestimates the experimental data as seen in Figures \ref{tp-18mev}, \ref{transf-56MeV} and \ref{ic+seq+cluster}.  In order to perform a more realistic cluster model calculation, a broader space for the intrinsic states of the two-neutron coupling with respect to the core was considered. This model is referred as microscopic cluster model, as recently shown by Carbone \textit{et al.} in Ref.\cite{CFC17}. The spectroscopic amplitudes for the microscopic cluster model were obtained from shell-model calculations using the Moshinsky transformation brackets \cite{Mosh59}. These transformations are made from individual ($j-j$) coupling to relative and center of mass coordinates ($LS$ coupling) for the harmonic oscillator wave functions of the two-particle system.

The number of possible combinations in this approach becomes quickly large and in Table \ref{cluster-micro} we only show the spectroscopic amplitudes used in the overlap of the $^{28}$Si ground state for the  $^{30}$Si  states. The spectroscopic amplitudes for $^{16}$O and $^{18}$O were obtained from Ref. \cite{CFC17}. 

\begin{figure}[H]
\graphicspath{{figuras/}}
\includegraphics[width=0.45\textwidth]{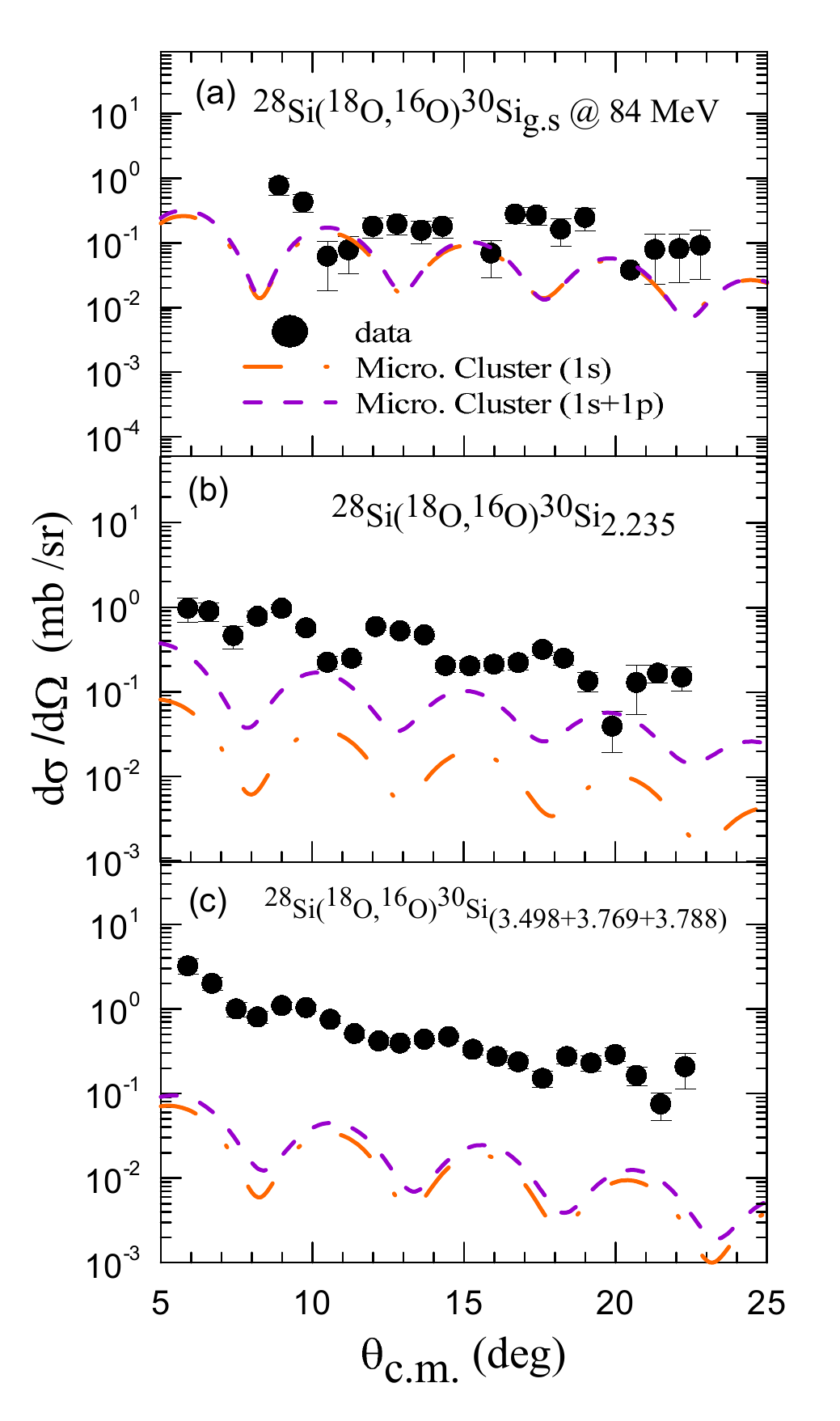}
\caption{Comparison of the experimental angular distribution with the microscopic cluster model calculations for the $^{28}$Si($^{18}$O,$^{16}$O)$^{30}$Si reaction at 84 MeV.}
\label{cluster-micro-fig}
 \end{figure}

The comparison between experimental data at 84 MeV and calculations performed within the microscopic cluster model is shown in Fig. \ref{cluster-micro-fig}. To assess the relevance of different intrinsic configurations of the two-neutron cluster, we consider that the two neutrons are coupled in the 1s ($\textnormal{n}=1$ e $\textnormal{l}=0$) and in the 1p ($\textnormal{n}=1$ e $\textnormal{l}=1$). One can observe that the inclusion of the 1p orbital affects only the cross sections of $^{30}$Si$_{2.235}(2^+)$. This means that the contribution of the p-orbital is important. Here, it is important to mention that, because of the parity conservation, given by orbital angular momentum transfer ($L$), some form factors were not considered. Even in the microscopic cluster model, the angular distributions for the first state (2.235 MeV) and the summed excited states the results are lower than the experimental data. Similarly as obtained recently for the $^{13}$C($^{18}$O,$^{16}$O)$^{15}$C reaction, the results of the microscopic cluster model are of the same order of magnitude as those obtained using the independent coordinates model.

\begin{table*} [ht]
\caption{Two-neutron spectroscopic amplitudes for CRC calculations obtained by shell model calculations with \textit{psdmod} interaction. n$_1$l$_1$j$_1$ n$_2$l$_2$j$_2$ are the principal quantum numbers, the orbital and the total angular momenta of the neutron 1 and 2 respect to the core,respectively, J$_{12}$ is the angular momentum of two neutron system, n,l,N,$\Lambda$ are the quantum numbers of the cluster wave function, L is the total orbital angular momentum, and S is the total spin of the two neutrons.}
\footnotesize
\centering
\begin{tabular}{c|c|c|c|c|c|c|c|c|c|c|c}
\hline \textbf{Initial State}&\textbf{nlj1j2} & \textbf{Final State} & \textbf{J12} & \textbf{Spect. Ampl.} &n & l& N& $\Lambda$ & L & S & \textbf{Spect. Ampl. (cm)} \\ \hline

                       & $(1p_{3/2})^2$  &                        &   &  0.007 &1 & 0& 2& 0& 0& 0& 0.009\\  
                       & $(1p_{1/2})^2$  &                        &   &  0.011 &&&&&&&\\ \cline{5-12} 
$^{28}$Si$_{g.s}(0_1^+)$ & $(1d_{5/2})^2$  & $^{30}$Si$_{g.s}(0_1^+)$ & 0 &  0.410 &&&&&&&\\  
                       & $(1d_{3/2})^2$  &                        &   &  0.576 &1 & 0& 3& 0& 0& 0& 0.515\\  
                       & $(2s_{1/2})^2$  &                        &   &  0.518 &&&&&&&\\ \hline

                       & $(1p_{3/2})^2$          &                        &   & -0.0003 &1 & 0& 1& 2& 2& 0&-0.003\\  
                       & $(1p_{3/2})(1p_{1/2})$  &                        &   & -0.007 &&&&&&&\\ \cline{5-12} 
$^{28}$Si$_{g.s}(0_1^+)$ & $(1d_{5/2})^2$      & $^{30}$Si$^*_{2.235}(2_1^+)$ & 2 & -0.077 &&&&&&&\\  
                       & $(1d_{5/2})(1d_{3/2})$  &                        &   &  0.026 &&&&&&&\\  
                      & $(1d_{5/2})(2s_{1/2})$  &                        &   &  0.011 &1 & 0& 2& 2& 2& 0& 0.082\\  \cline{5-12}
                       


                       & $(1d_{5/2})^2$          &                        &   & -0.076 &&&&&&&\\  
                       & $(1d_{5/2})(1d_{3/2})$  &                        &   &  0.026 &&&&&&&\\  
                       & $(1d_{5/2})(2s_{1/2})$  &                        &   &  0.011 &1&1&1&3&2&1& 0.171\\  
                       & $(1d_{3/2})^2$          &                        &   & -0.248 &&&&&&&\\ 
                       & $(1d_{3/2})(2s_{1/2})$  &                        &   & -0.537 &&&&&&&\\ \cline{5-12}


                       & $(1d_{5/2})^2$          &                        &   & -0.076 &&&&&&&\\  
                       & $(1d_{5/2})(1d_{3/2})$  &                        &   &  0.026 &&&&&&&\\  
                       & $(1d_{5/2})(2s_{1/2})$  &                        &   &  0.011 &1&1&2&1&2&1& -0.112\\  
                       & $(1d_{3/2})^2$          &                        &   & -0.248 &&&&&&&\\ 
                       & $(1d_{3/2})(2s_{1/2})$  &                        &   & -0.537 &&&&&&&\\ \hline
                        
                       & $(1p_{3/2})^2$          &                        &   &  0.004 &1&0&1&2&2&0& 0.0003\\  
                       & $(1p_{3/2})(1p_{1/2})$  &                        &   & -0.005 &&&&&&&\\ \cline{5-12}
$^{28}$Si$_{g.s}(0_1^+)$ & $(1d_{5/2})^2$      & $^{30}$Si$^*_{3.498}(2_2^+)$ & 2 &  0.148 &&&&&&&\\ 
                       & $(1d_{5/2})(1d_{3/2})$  &                        &   &  0.013 &&&&&&&\\  
                       & $(1d_{5/2})(2s_{1/2})$  &                        &   &  0.091 &1&0&2&2&2&0& 0.053\\  
                       & $(1d_{3/2})^2$          &                        &   &  0.232 &&&&&&&\\  
                       & $(1d_{3/2})(2s_{1/2})$  &                        &   &  0.167 &&&&&&&\\ \cline{5-12}
                       


                       & $(1d_{5/2})(1d_{3/2})$  &                        &   &  0.013 &&&&&&&\\  
                       & $(1d_{5/2})(2s_{1/2})$  &                        &   &  0.091 &1&1&1&3&2&1& -0.078\\  
                       & $(1d_{3/2})^2$          &                        &   &  0.232 &&&&&&&\\  
                       & $(1d_{3/2})(2s_{1/2})$  &                        &   &  0.167 &&&&&&&\\ \cline{5-12}


                       & $(1d_{5/2})(1d_{3/2})$  &                        &   &  0.013 &&&&&&&\\  
                       & $(1d_{5/2})(2s_{1/2})$  &                        &   &  0.091 &1&1&2&1&2&1& 0.051\\  
                       & $(1d_{3/2})^2$          &                        &   &  0.232 &&&&&&&\\  
                       & $(1d_{3/2})(2s_{1/2})$  &                        &   &  0.167 &&&&&&&\\ \hline

                       
$^{28}$Si$_{g.s}(0_1^+)$ &$(1d_{5/2})(1d_{3/2})$ & $^{30}$Si$^*_{3.769}(1_1^+)$ & 1  & -0.027 &1&1&1&3&2&1& 0.249\\  
                       & $(1d_{3/2})(2s_{1/2})$  &                        &   &  -0.594 &&&&&&&\\ \cline{5-12}  
                       
                       &$(1d_{5/2})(1d_{3/2})$   &                        &   & -0.027 &1&1&2&1&2&1& -0.163\\  
                       & $(1d_{3/2})(2s_{1/2})$  &                        &   &  -0.594 &&&&&&&\\ \hline 
                       
                       & $(1p_{3/2})^2$  &                        &   &  -0.003 &1&0&2&0&0&0& 0.003\\  
                       & $(1p_{1/2})^2$  &                        &   &  0.013 &&&&&&&\\ \cline{5-12} 
$^{28}$Si$_{g.s}(0_1^+)$ & $(1d_{5/2})^2$  & $^{30}$Si$_{3.788}(0_2^+)$ & 0 &  0.041 &&&&&&&\\   
                       & $(1d_{3/2})^2$  &                        &   &  0.528 &1&0&3&0&0&0& -0.046\\  
                       & $(2s_{1/2})^2$  &                        &   &  -0.427 &&&&&&&\\ \hline  
                        
                       

\end{tabular}
\label{cluster-micro}
\end{table*}

\section{\label{conc}Conclusions}
  
In the present work, we have analyzed the experimental data obtained for the two-neutron transfer in the $^{28}$Si(t,p)$^{30}$Si reaction at 18 MeV and the $^{28}$Si($^{18}$O, $^{16}$O)$^{30}$Si reaction at 56 MeV and new data at 84 MeV. 


Calculations of simultaneous transfer were performed within the CRC approach considering the extreme cluster and independent coordinates models to describe the two-neutron system. The microscopic cluster model was considered as well, to assess the contribution of the 1p component of the two-neutron cluster. The addition of 1p orbital in the calculation affected only the state at 2.235 MeV ($2^{+}$). The sequential transfer was calculated within the two-step CCBA. 

All the results for the cross section of two-neutron transfer at 84 MeV presents a good description of the period of oscillation although a phase difference around three degrees with respect to the experimental data is present. The extreme cluster model overestimates the experimental data in almost all angular distributions studied here. Calculations for the two-neutron transfer leading to the ground state ($0^{+}$) of the $^{30}$Si show that very similar angular distributions are obtained considering the simultaneous, in the independent coordinate scheme, and sequential mechanisms in both (t,p) and ($^{18}$O,$^{16}$O) reactions. Instead, the simultaneous transfer is dominant for the population of the ground state in $^{14}\textnormal{C}$ and $^{18}\textnormal{O}$. Here, in the case of $^{30}\textnormal{Si}$, the collective nature of nuclei interfere with the short-range correlation between the two neutrons and the sequential transfer becomes a relevant mechanism. Similar conclusion has been obtained in the analysis of the two-neutron transfer leading to the $2_{1}^{+}$ state in $^{66}\textnormal{Ni}$ \cite{PSM17}. Calculations performed at different bombarding energies exhibit similar results. Based on our results we conclude that the interplay between simultaneous and sequential two-neutron transfer is intrinsically related to the nuclear structure of the target nuclei.

\section*{Acknowledgment}
This project has received funding from the European Research Council (ERC) under the European Union’s Horizon 2020 research and innovation programme (grant agreement No 714625).The Brazilian authors acknowledgment partial financial support from CNPq, FAPERJ and CAPES and from INCT-FNA (Instituto Nacional de Ci\^ {e}ncia e Tecnologia- F\' isica Nuclear e Aplica\c {c}\~ {o}es).  

\pagebreak

\end{document}